%% file: main.tex
\documentclass[fleqn,10pt]{wlscirep}
\usepackage[utf8]{inputenc}
\usepackage[T1]{fontenc}
\usepackage{lscape}

\usepackage{rotating}
\usepackage{graphicx}
\usepackage{subfig}

\title{Price Formation in Field Prediction Markets: the Wisdom in the Crowd}

\author[1,*]{Frederik Bossaerts}
\author[2]{Nitin Yadav}
\author[2,3]{Peter Bossaerts}
\author[1]{Chad Nash}
\author[1]{Torquil Todd}
\author[1]{Torsten Rudolf}
\author[1]{Rowena Hutchins}
\author[4]{Anne-Louise Ponsonby}
\author[1]{Karl Mattingly}

\affil[1]{Dysrupt Labs}
\affil[2]{Brain, Mind, and Markets Lab, University of Melbourne}
\affil[3]{Cambridge University}
\affil[4]{The Florey Research Institute of Neuroscience and Mental Health}
\affil[*]{Corresponding Author: frederikbossaerts@dysruptlabs.com}


\begin{abstract}
Prediction markets are a popular, prominent, and successful structure for a collective intelligence platform. However the exact mechanism by which information known to the participating traders is incorporated into the market price is unknown. Kyle (1985) detailed a model for price formation in continuous auctions with information distributed heterogeneously amongst market participants. This paper demonstrates a novel method derived from the Kyle model applied to data from a field experiment prediction market. The method is able to identify traders whose trades have price impact that adds a significant amount of information to the market price. Traders who are not identified as informed in aggregate have price impact consistent with noise trading. Results are reproduced on other prediction market datasets. Ultimately the results provide strong evidence in favor of the Kyle model in a field market setting, and highlight an under-discussed advantage of prediction markets over alternative group forecasting mechanisms: that the operator of the market does not need to have information on the distribution of information amongst participating traders.
\end{abstract}

\begin{document}

\flushbottom
\maketitle


\input{S_introduction}
\input{S_marketdesign}
\input{S_results}
\input{S_discussion}

\bibliography{bibliography}


\subsection{Acknowledgements}
We thank the University of Melbourne's Brain, Mind, and Markets lab for their expertise and feedback on the work presented in this manuscript. The research and development presented in this paper was partly funded by the Australian Government through the government's R\&D tax offset programme.

\subsection{Author Contributions}
FB designed and performed the analysis presented, and wrote the manuscript. CN, TT, and TR built the market platform which created the data, and provided feedback on the analysis. RH and KM operated the platform, including generating and resolving market contracts over the lifetime the data cover. PB, NY, ALP provided feedback and guidance in analysis and presentation of results.

\subsection{Data Availability}
The Almanis dataset analysed in this study are available from Dysrupt Labs, but restrictions apply to the availability of these data, which were used under license for the current study, and so are not publicly available. Data are however available from the authors upon reasonable request and with permission of Dysrupt Labs.

The replication data set analysed during the current study is publicly available through the R package at:\linebreak https://github.com/MichaelbGordon/PooledMarketR

\pagebreak
\section{Supporting Information Appendix (SI)}
\input{SOM_LMSR}

\input{SOM_Methods}
\input{SOM_ROC}

\input{SOM_TLS}
\input{SOM_Truncation}

\end{document}

%% file: S_introduction.tex
\section{Introduction}
Accurately predicting the future in order to properly prepare for epidemics, financial crashes and referendum debacles is paramount for success. Prediction markets are a promising mechanism that utilizes the supposed “wisdom of the crowd”, aggregating the knowledge of a diverse group of individuals into a precise forecast, often beating predictions from experts.\cite{Arrow:2008km} Recently, the the Commodity Futures Trading Commission (CFTC) reversed its decades long stance on event contracts and allowed for regulated real-money prediction market exchanges to be set up in the US.\cite{cftc_2020}

However, the exact mechanism by which prediction market prices come to reflect the information known to the participating traders is unknown. The most common explanatory mechanism is that prediction market prices become informed because they average many pieces of information dispersed among the participating traders. Random noise in individual trader forecasts (beliefs) is macroscopically cancelled out, reducing mispricing and providing a better forecast than any of the individuals alone. A direct consequence of this view is that market prices should become better according to the total number of participating traders, obeying a law of large numbers.

This view is supported by Plott and Sunder \cite{plott1988rational}, which found that under simple circumstances and complete markets with evenly distributed heterogeneous information, this information in aggregate is reflected in the prices (confirming rational expectations theory). However, recent evidence from laboratory experiments has cast doubt on the ability of markets to aggregate evenly distributed heterogeneous information under different conditions.\cite{Page&Siemroth:2018,DeSantis:2019} Instead, experiments show that better pricing depends on there being more individual traders in the market who are either better informed\cite{Page&Siemroth:2017,Bossaerts:2014ed} or better at analyzing homogeneous information\cite{Asparouhova:2015dx} - they have an ‘edge’ over the other traders in the market. These findings do not necessarily contradict those of Plott and Sunder, instead they shed light on the lack of generalizability of the information aggregation hypothesis to more complex experimental environments; thereby casting doubt that the underlying mechanism at play is a simple averaging of information known to participating traders.

In field markets, there is little evidence towards any mechanism of price formation. This is because unlike in laboratory experiments, the distribution of information relevant to the pricing of securities in a given market is not observable (or controllable). As such, whether the information relevant to a given market is dispersed amongst market participants evenly and homogeneously, evenly and heterogeneously, unevenly and heterogeneously, etc is not known. This poses extreme difficultly to ascertain how the information reflected by the prices that are observed relates to the information held by the market participants. Consequently one cannot directly identify which mechanism is at play: whether the information reflected in the prices is an average of the information held by all participants or the amplification of superior information in the hands of a few traders (this superior information being acquired either through access to non-public information or a superior processing of publicly available information as in Asparouhava et al.\cite{Asparouhova:2015dx}).

Kyle (1985)\cite{Kyle:1985wu} put forward a model of price formation in continuous auctions that describes the market as a game between three types of players: informed traders (who have access to private information on the liquidation value of the traded securities), noise traders (who have no information and trade randomly), and the market maker (who sets prices based on order flow). The model assumes that the informed traders are risk neutral and seek to maximize their expected profit, and details the optimal strategy for the informed trader as a function of their information edge relative to the current market prices and the market depth - amount of order flow from the noise traders.\cite{bernhardt2002intraday}

The optimal strategy of the informed trader is to purchase the securities that are mispriced (priced either too low or too high based on the informed trader’s private information), and to do this when there is a large amount of noise trader order flow to “camouflage” their activity.\cite{Kyle:1985wu, HOLDEN:2012kr} This results in the informed trader manifesting price-sensitive behavior - they alter their demand for the traded security as the price offered by the market changes. Their price sensitivity will always be negative (reverting) - their demand will always point in the direction of their private information, in the opposite direction of the price vector from that information. This is in contrast to positive price-sensitivity, which is when demand responds to a change in prices by increasing along the price vector (momentum) (For the remainder of this paper we will use “price-sensitive” in the former sense, rather than the latter). This feature allows one to identify them, as was done in Asparouhova et al. (2015).\cite{Asparouhova:2015dx}

The method put forward in this paper aims to identify traders within each market who display price-sensitive behavior, and examines the informativeness of their trades and the impact on the overall market in which they take place. In short, under the Kyle model informed traders should be price sensitive, and we investigate if traders who demonstrate price-sensitivity in a real world market are informed. By doing so we aim to understand whether prices become good forecasts of future events because of the wisdom of the crowd or the wisdom in the crowd.

%% file: S_marketdesign.tex
\section{Market Design}
The primary market studied here was designed to extract information from a large crowd of participants from across the world interested in geopolitical and financial events. It was constructed utilizing the Logarithmic Market Scoring Rule (LMSR) market maker.\cite{Hanson:2012fh} The LMSR is widely used, including at Microsoft, Yahoo!, Inkling Markets, Consensus Point, Carnegie Mellon,\cite{Othman:2013ih} and across various academic studies.\cite{dreber2015using, camerer2016evaluating, camerer2018evaluating, forsell2019predicting}

LMSR markets are a form of subsidized market where the market maker supplies and demands “Arrow-Debreu'' shares according to a fixed cost function. The price of each share $p_i$ corresponding to asset $i$ is adjusted according to the quantity $q$ of each of the $N$ shares in the market according to Equation \ref{eq:lmsr_price}. When the market is “settled” and the shares expire, the shares are liquidated according to the world state specified by the market operator. Shares corresponding to the correct world state pay off \$1, and all others pay \$0. Participating traders purchase shares directly from the market maker, who adjusts prices according to this fixed cost function. Within an LMSR market, the price impact of individual traders is governed by a “liquidity factor” and the cash endowed to the traders.\cite{Hanson:2012fh} In the markets studied here, traders were endowed with 1000 points, and the LMSR liquidity factor $B$ was set at 150. These values were chosen so that traders had the ability to push the market price to match their perceived likelihood of an event. LMSR markets are thus quote driven markets (another example of quote driven markets are foreign currency markets). 

\begin{equation} \label{eq:lmsr_price}
	p_{i}(q) = \frac{e^\frac{q_i}{B}}{\sum_{j=1}^{N}e^\frac{q_j}{B}}
\end{equation}

The field experiment took 1 1/2 years. Participants were endowed with a point balance and paid the real money incentives based on their net improvement of this balance at regular intervals. Participants’ incentives were substantial, amounting to approximately 70,000 GBP (\$105,000 USD at the time of the field experiment). 

%% file: S_results.tex
\section{Results}

In the model of Kyle (1985)\cite{Kyle:1985wu}, an informed trader has private information on the true liquidation value of the security being traded. In the lab experiment of Asparouhova et al. (2015)\cite{Asparouhova:2015dx} it was demonstrated that price quality improved with the number of traders who traded in the direction of the correct liquidation value. This liquidation value and its correct expectation were controlled, and hence known to the experimenters. However in our field experiment, we do not know what the correct expected liquidation value is, even after a market has settled and been liquidated. This is because, even if a market settles as “True”, this does not mean that this was drawn from a distribution of almost-surely True ($P(True) = 1$). The lack of knowledge of the correct expected liquidation value makes this setting unlike prior laboratory experiments such as \cite{Asparouhova:2015dx}). 

Here we mention that the LMSR market maker itself is not exactly the market maker of Kyle (1985) \cite{Kyle:1985wu}. The LMSR market maker is not Bayesian, and it is a net subsidizer rather than a zero-profit actor. The LMSR offers at each moment in time the price that the previous trader pushed it to based on that trader's trade, treating orders from all traders identically. Therefore it acts as a hybrid of a Kyle model market maker and noise trader. The subsidy of the LMSR is, unlike the loss of the noise traders in Kyle's model, bounded by its pricing rule and so its maximum subsidy is always finite\cite{Hanson:2012fh}.

We propose a novel method to detect price-sensitive traders. This method is as follows: we regress the price impact of a trader’s trade onto the difference between the market price before the trade and the price the same trader pushed the market to on their previous trade:

\begin{equation} \label{eq:price_sense_estimation}
	\Delta p_{i}(t_i+1)=\alpha_i+\beta_i (p_0 (t_i+1)-p_{m,i}(t_i)) + \epsilon(t_i)
\end{equation}

where $t_i$ is the integer business time for trader $i$ (and thus refers to the $t_i$th trade of trader $i$), $\Delta p_{i}(t_i+1) = p_{m,i}(t_i+1) - p_0(t_i+1)$ is the price impact of trader $i$’s trade on the price of the asset at time $t_i+1$, $p_0(t_i+1)$ is the market offer price of the asset at time $t_i+1$ before trader $i$’s order is executed, and $p_{m,i}(t_i)$ is the final marginal price trader $i$ traded to previously ($p_{m,i}(t_i+1)$ is therefore the final marginal price of trader $i$’s trade at time $t_i+1$ ). Note that between time $t_i+1$ and $t_i$ an arbitrary amount of "real-time" may have passed.

The regression in Equation \ref{eq:price_sense_estimation} is applied to every trader within a particular market, and those with significant negative $\beta_i$ ($t statistic < -1.65$) are deemed to be price-sensitive. Those with significant positive t-statistics are also technically price-sensitive in that their demand changes with changes in the market price, however this is not the kind of price-sensitivity that is optimal for an informed trader according to the Kyle framework. Figures \ref{fig:psense_example_trump} \& \ref{fig:psense_example_french} show examples of an individual price-sensitive trader’s trade history in the context of a larger market price history. Figures \ref{fig:npsense_example_trump} \& \ref{fig:npsense_example_brexit} show examples of non-price-sensitive traders.

\input{Figures/fig1}

Conceptually the method works because $p_{m,i}(t_i)$ provides an estimate of the trader’s subjective belief: it is the price at which the trader stopped trading the last time they traded. The offer price of the market before the subsequent trade, $p_0(t_i+1)$, provides an estimate of the market’s belief, and the difference between the two,  $p_0(t_i+1) - p_{m,i}(t_i)$, measures the approximate incentives for the trader to trade, provided their beliefs have not changed since their last trade. We hereby exploit a unique feature of a dealer/specialist system (the class of market makers to which the LMSR belongs), which is that every individual trade has price impact. A myopic, risk neutral trader will stop trading (and impacting the price) when the final price is aligned with their beliefs. The assumption that traders are approximately risk neutral is in line with Kyle’s theory, which also assumes risk neutrality of the informed traders. If traders are not approximately risk neutral, this regression is misspecified in a particular way (more on this later).

Critically, our method leaves open the possibility that the trader’s prior is wrong, and therefore those that demonstrate price-sensitive behavior are not by construction guaranteed to be better informed than those who do not. Even if informed traders as described by Kyle exist, they may not constitute a significant enough portion of the set of price-sensitive traders to allow this method to separate informed from uninformed.

If those who trade in a price-sensitive fashion are truly informed, the final marginal price of their trades $p_{m,i}(t_i)$ will be more informative than the price offered by the LMSR market maker at the moment they began their trade $p_0(t_i)$. Traders who are not trading in a price-sensitive manner provide a useful benchmark against which to gauge the added information value of the price-sensitive traders through their price impact.

We proceeded to test the hypothesis that the instantaneous price impact of the trades by price-sensitive traders added information on settlement direction of the market. Addition of information is defined here as an increase in the area under curve (AUC) between the LMSR offer price $p_0(t_i)$ and the final marginal price of the trade $p_{m,i}(t_i)$, with respect to the final settlement value of the market security (1 or 0). The AUC is a measure of the ordinal well-orderedness of the prices: if we randomly select a price from the set of markets that settled at 0, we should expect with high probability that this price is lower than another price randomly drawn from the markets that settled at 1 ; the AUC is this probability. Therefore an AUC > 0.5 indicates a better-than-random ordering of prices with respect to the settlement value of the market \cite{SOM}.

We split the trading data into groups: trades by price-sensitive traders, and trades by non-price-sensitive traders. For each group two ROC curves were computed: one for the initial market offer price $p_0(t_i)$ accepted at the beginning of each trade with respect to the final settlement value S of the market; and one for the final marginal price of each trade $p_{m,i}(t_i)$ with respect to the final settlement value S of the market. This procedure was repeated using subsets of trades from each group, stratified based on minimum price impact. Price impact was measured by the KL divergence between the initial offer price $p_0(t_i)$ and the trader’s final marginal price $p_{m,i}(t_i)$. This is to evaluate the effect of larger magnitude trades  - trades whose beginning and final prices are very different - on the ordinal structure of the prices. The filtration based on KL divergence is done in a cumulative, and not a binned fashion, to avoid the assumption that the derivative of the change in AUC with respect to the KL divergence exists. Figures \ref{fig:npsense_roc_1} - \ref{fig:psense_roc_3} shows these ROC curves for different subsets of both price-sensitive and non-price-sensitive trades.

The change in AUC from initial offer price $p_0(t_i)$ to final marginal price $p_{m,i}(t_i)$ is plotted against minimal price impact in figure \ref{fig:pimpact_curve}. Curves differentiate price-sensitive (red) and price-insensitive (blue) traders. Trades of price-sensitive traders with higher minimum price impact generate a higher gain in AUC; the opposite is true for the remaining traders. The latter even cause prices to move to levels with reduced predictive power. This is consistent with Kyle’s theory on informed traders, which models non-informed traders as noise traders.\cite{Kyle:1985wu} A random trade against an informed market price would, on average, reduce the AUC of the final price. An increase in the minimum KL-divergence of a noise trader would correspond to an increase in the variance of their noise, leading to a more severe deterioration of the information present in the market price post-trade. This is indeed what is seen. Therefore the traders detected as price-sensitive must be reliably trading in an informed fashion with respect to the prices offered at the time of their trade, and those classified as non-price sensitive in aggregate trade in a noisy fashion.

\input{Figures/fig2}

The above result was replicated on a dataset compiled from four other LMSR prediction market studies: Dreber et al 2015 (PNAS), Camerer et al 2016 (Science), Camerer et al 2018 (Nat Human Behav), Forsell et al 2019 (Fig \ref{fig:reproduction_pimpact})\cite{dreber2015using, camerer2016evaluating, camerer2018evaluating, forsell2019predicting}. These studies aimed to predict experimental replication using a crowd of domain experts. We note that the method and analysis thus far discussed in this paper was developed entirely prior to accessing these external datasets.

The configuration of the market maker and the endowments of the participating traders in these studies differed from the Almanis setup in that traders faced much tighter budget constraints in view of the liquidity factor of the LMSR (Fig \ref{fig:market_configuration_comparison}). This can easily be seen empirically in Fig \ref{fig:reproduction_pimpact} as the range of price impacts is substantially smaller in terms of cost as a fraction of endowment than on the Almanis data - the limits representing the same percentile of trades. The stronger budget constraints cause traders not to push prices as strongly towards levels that reflect their beliefs. Instead, price changes reflect only a fraction of their disagreement with the market, an effect that is not unlike that caused by extreme risk aversion. Ensuing price adjustments become a noisy estimate of the difference between a trader’s belief and market beliefs, inducing errors in variables in our method when estimated via ordinary least squares. We emphasize that this error in variables is different from the traditional one encountered in econometrics since it is not treatable through the common instrumental variables approach. Instead we use the Total Least Squares estimation method (for more information see supplementary materials). \cite{golub1996matrix, markovsky2007overview,SOM}

Persistence of price-sensitive traders across markets was examined. The distribution of the number of markets in which a trader was price-sensitive is highly skewed: 75\% of traders are price-sensitive in fewer markets than the mean. Traders who appear as price-sensitive in a number of markets are not price-sensitive in every market in which they participate. For example, the maximal trader, who was price-sensitive in 64 markets, was also price-insensitive in 42 other markets. Only one trader was price-sensitive with 100\% frequency, but this trader only participated in one market. Therefore, it is not obvious that evidence of price-sensitivity in a market provides much information about the likelihood of price-sensitivity in a future market. This can be explained intuitively as a result of non-transitivity of an informational advantage from one market (forecasting topic) to another. The Almanis marketplace covered a wide variety of topics, from geopolitical to macroeconomic to environmental to epidemiological, and therefore this lack of transitivity is perhaps expected.

However, if we examine the experimental replication markets, the setup is different - the topics of the markets are all concentrated on the domain expertise of market participants. Therefore, by utilizing the price-sensitivity of a trader in one market as an indicator of informedness across all markets in which they participated, we are able to further strengthen the results found by gaining better separation of the price-sensitive and non-price-sensitive price impact curves (Fig \ref{fig:reproduction_pipact_transitive}).

\input{Figures/fig3}

Kyle’s model of price formation states that in the simple case of a single informed trader, that trader acts as a monopolist and is able to extract substantial profit from the market. However, as the number of informed traders increases, they must compete with one another in markets with finite depth, leading to reduced profits. The consequence on the market price of this competition is that the price converges more rapidly to the information of the informed traders.\cite{HOLDEN:2012kr} Consequently, markets in which there are more informed traders should converge more quickly to the superior information of said traders.

In order to explore the effect of the number of detected price-sensitive traders on the convergence of forecasting power of the market price, markets needed to be selected for periods when price-formation was occurring. A significant number of raw market time-series had a long ”tail” of low trade activity before settlement due to the outcome of the market being near-certain long before the official settlement time of the market. Therefore market time-series were truncated before the analysis to remove periods of low trade activity. The methodology of truncation was as follows: if the overall trade frequency of the market fell below two trades per 24 hours, the trade closest to the settlement was dropped. This process was repeated until the market either attained the desired average trade frequency or dropped below 25 total trades, in which case the market was dropped from the analysis. No further selection criteria were applied to the market data. The analysis of price-sensitivity and market forecasting performance was performed on the truncated data.

The markets were grouped by the number of price-sensitive traders detected in the truncated version of the market time-series. Of the 152 truncated markets analyzed, 14.5\% had no price-sensitive traders, 19.1\% had 1 price-sensitive trader, 21.2\% had 2 price sensitive traders, 13.8\% had 3 price-sensitive traders, 11.2\% had 4 price-sensitive traders, and 20.2\% had 5 or more price-sensitive traders. The mean number of price-sensitive traders in a market was 3.06 (std. dev. = 3.47), and the mean proportion of price-sensitive traders to total traders in the market was 7.3\% (Interquartile Range (IQR) 4.06\% to 10.0\%). Since a trader needs to have at least 3 trades in order to compute a t-statistic, a relevant quantity when discussing detection error is the proportion of price-sensitive traders to the number of traders with computed t-statistics. The distribution of the number of price-sensitive traders to traders with three or more trades had a median of 33.3\% (IQR 20.0\% to 44.9\%).

The final prices of these markets were recorded every day prior to the end of the active period (EAP). For each market group an ROC curve was constructed from the prices N days prior to EAP (for each N from 1 to 14) (Fig \ref{fig:market_price_convergence}). The time evolution of the AUC is shown in Fig \ref{fig:market_price_convergence}. A temporally averaged ROC curve was produced for each group using a threshold method \cite{Fawcett:2006gr}. In order to estimate errors for the AUC, a bootstrapping method was used (Fig \ref{fig:market_price_convergence}). Both the 2-3 and 4+ price-sensitive trader markets have larger AUCs than 1 or 0 price-sensitive trader markets. The difference in AUCs is significant in two sample z-tests under bootstrapped errors (p < 0.001 for all). 

As can be seen across Figure \ref{fig:market_price_convergence}, markets with more price-sensitive traders have prices that are more informative and more informative earlier (with respect to the truncated time-series), of the final settled value of the market, compared to markets with fewer price-sensitive traders.

\input{Figures/fig4}

These results corroborate that, in the Kyle model, the presence of competing insiders causes prices to reflect valuable information more quickly. Experimentally this phenomenon has also been shown to exist \cite{Bossaerts:2014ed}. However, it is worth remarking that even if the Kyle model is assumed to be the correct model of price-formation, it is not true that one would be able to reproduce the above result in a field market. The reason being that the in field market the number of informed traders is not a directly observable quantity. Our method can reproduce Kyle model implications in the field, however critically it relies on there being sufficient trading opportunities in each market for an informed individual to trade a sufficient number of times, at sufficient magnitude. If a market converges too quickly to the informed price (for example: there are relatively few noise traders compared to informed traders), then the informed traders’ price sensitivity will not be estimable, and it will appear as if there are no price-sensitive traders in the market. Yet market prices will be very good. So a lack of detected price-sensitive traders can mean either that there are no informed traders, or that there are too many. This may also mean that the data set of markets studied must be such that the markets with 0 or 1 Price sensitive traders are markets in which there may in truth be no informed traders - no one is informed, or the information edge of the informed is very small with respect to a random guess.

On the experimental replication market data sets, the result of Figure \ref{fig:market_price_convergence} does not reproduce for precisely this reason. Markets in which price-sensitive traders appear are markets that have many opportunities for these traders to act on their information. Markets where they do not appear are largely dominated by lack of opportunity because prices are too good, rather than the participants not being informed. Combined with the aforementioned budget constraint issues, the result is that the number of price-sensitive individuals detected actually inversely relates to the informativeness of the market price.

The above is also why the best-performing markets in terms of price-quality on Almanis appear to be markets with 2-3 price sensitive traders: in order to support a large number of informed traders manifesting price-sensitivity (4+), the pricing of the market must be more reliably poor.

%% file: Figures/fig1.tex
\begin{figure*}[ht]
    \subfloat[]{
        \includegraphics[width=0.48\textwidth]{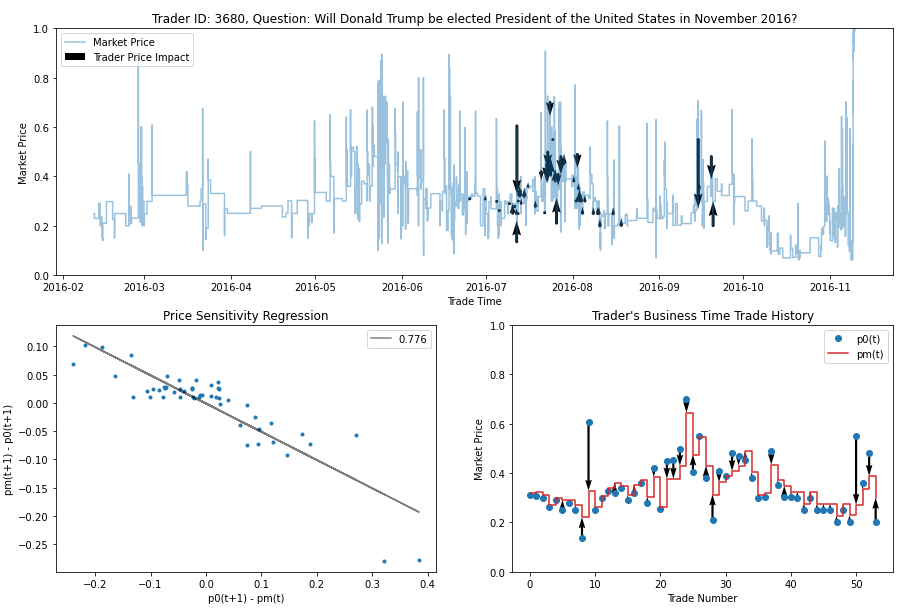}
        \label{fig:psense_example_trump}
    }
    \hfill
    \subfloat[]{
        \includegraphics[width=0.48\textwidth]{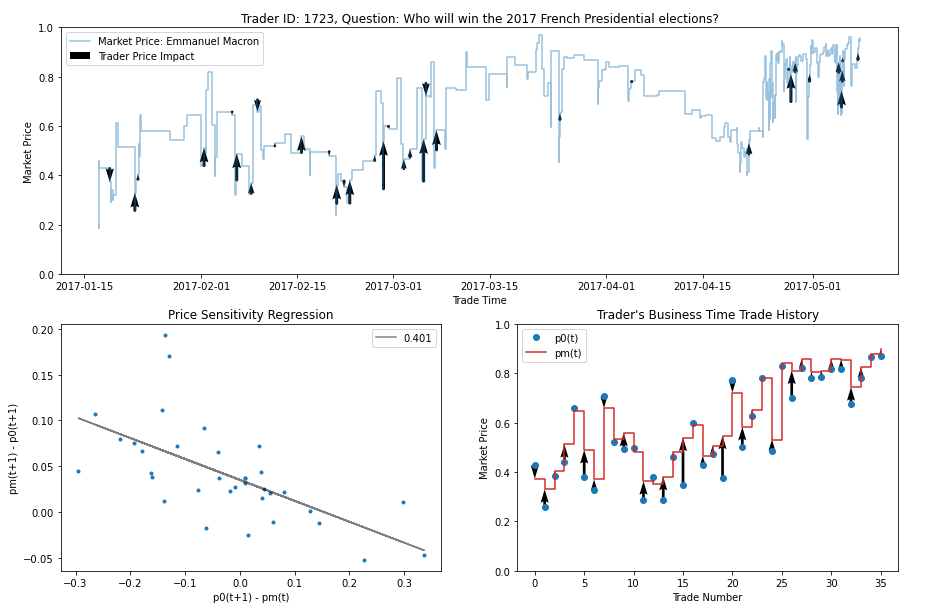}
		\label{fig:psense_example_french}
	}
	\vfill
    \subfloat[]{
        \includegraphics[width=0.48\textwidth]{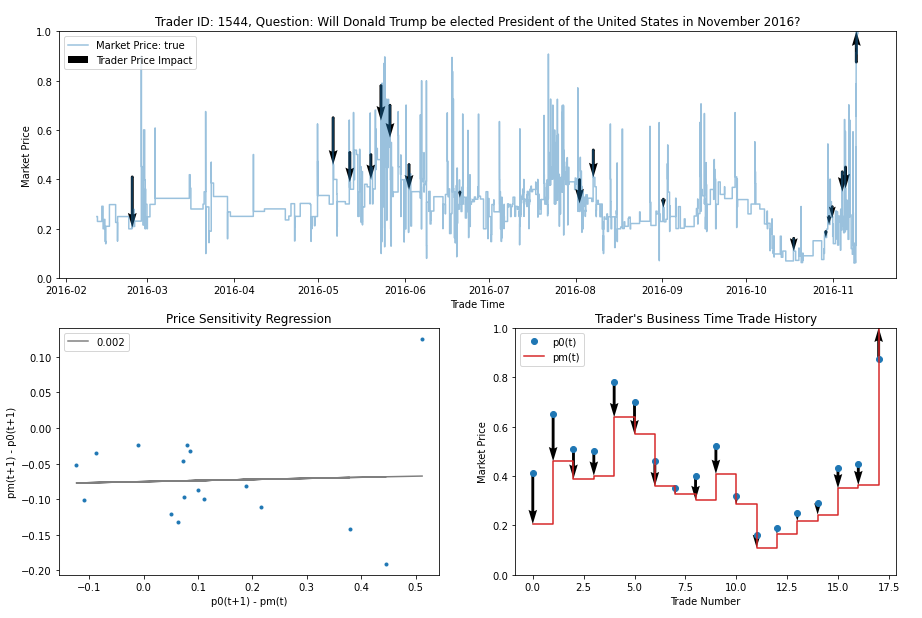}
        \label{fig:npsense_example_trump}
	}
	\hfill
	\subfloat[]{
	    \includegraphics[width=0.48\textwidth]{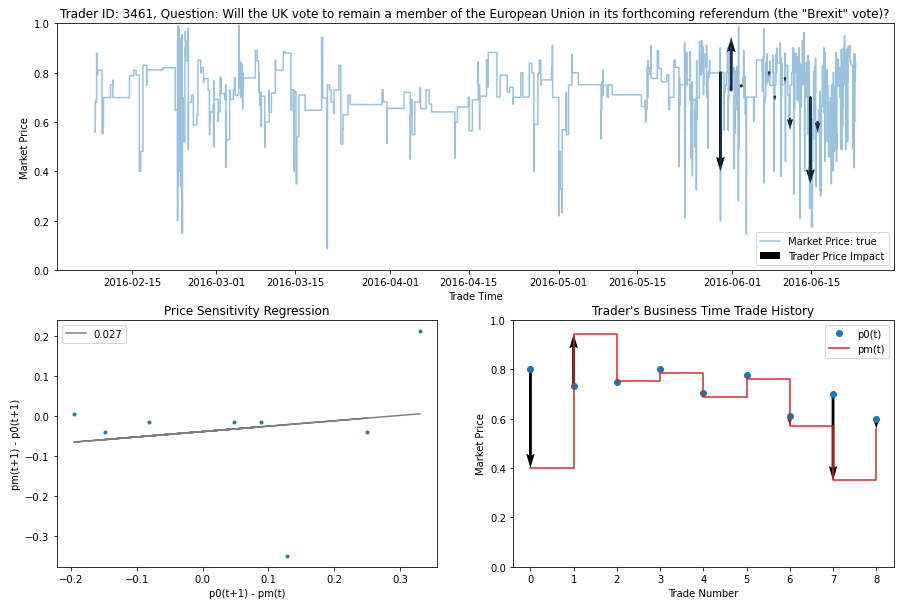}
	    \label{fig:npsense_example_brexit}
	}
\caption{\protect \subref{fig:psense_example_trump} \& \protect \subref{fig:psense_example_french}: Price Sensitive Trader examples, (i) shows the overall market time series (blue) with the trader's trades and price impact shown by the black arrows, (ii) (bottom left) the corresponding price-sensitivity regression (legend displays $R^2$), (iii) (bottom right) trader's transaction history in the trader's business time $t_i$, with market offered prices $p_0(t_i)$ (blue) and final price due to trader's price impact $p_{m,i}(t_i)$ (red). \protect \subref{fig:npsense_example_trump} \& \protect \subref{fig:npsense_example_brexit}: Non Price Sensitive Trader examples. (i)-(iii) are the same as before.}
\label{fig:examples_figure}
\end{figure*}

%% file: Figures/fig2.tex
\begin{figure*}
\begin{minipage}{.26\textwidth}
    \subfloat[]{
			\includegraphics[width=\textwidth]{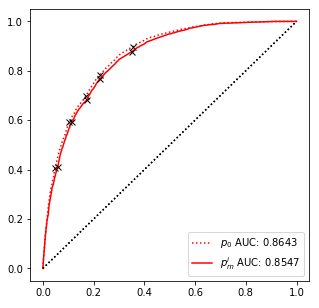}
			\label{fig:npsense_roc_1}
	}
	\vfill
	\subfloat[]{
			\includegraphics[width=\textwidth]{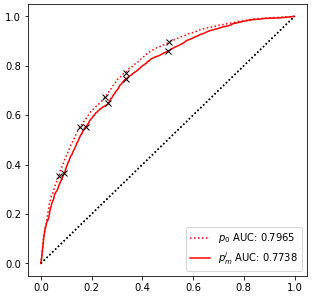}
			\label{fig:npsense_roc_2}
	}
	\vfill
	\subfloat[]{
			\includegraphics[width=\textwidth]{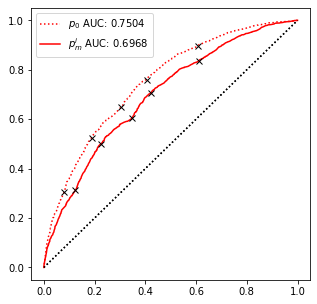}
			\label{fig:npsense_roc_3}
	}
\end{minipage}
\begin{minipage}{.26\textwidth}
    \subfloat[]{
			\includegraphics[width=\textwidth]{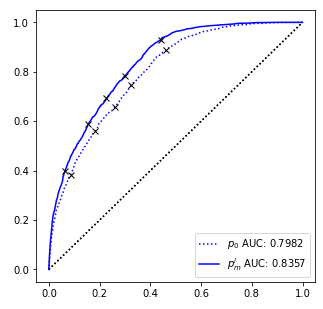}
			\label{fig:psense_roc_1}
	}
	\vfill
	\subfloat[]{
			\includegraphics[width=\textwidth]{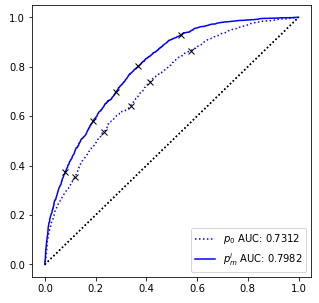}
			\label{fig:psense_roc_2}
	}
	\vfill
	\subfloat[]{
			\includegraphics[width=\textwidth]{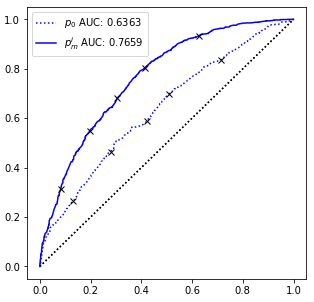}
			\label{fig:psense_roc_3}
	}
\end{minipage}
\begin{minipage}{.48\textwidth}
  \subfloat[]{
    \centering
    	\includegraphics[width=\textwidth]{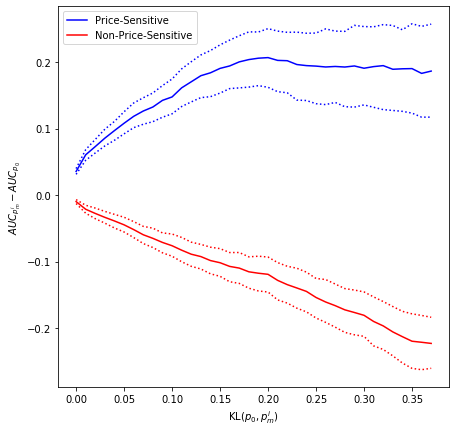}
			\label{fig:pimpact_curve}
    }
\end{minipage}
\caption{ROC curves of: Non-price-sensitive trades offer $p_0$ and marginal $p_{m,i}$ prices, filtered by minimum price impact (in KL divergence) of \protect \subref{fig:npsense_roc_1} 0.0, \protect \subref{fig:npsense_roc_2} 0.0524, \protect \subref{fig:npsense_roc_3} 0.0094 ; Price-sensitive trades offer $p_0$ and marginal $p_{m,i}$ prices, filtered by minimum price impact (in KL divergence) of \protect \subref{fig:psense_roc_1} 0.0, \protect \subref{fig:psense_roc_2} 0.0524,\protect \subref{fig:psense_roc_3} 0.0094 (the price impact cutoffs correspond to the 0.0, 0.5, and 0.75 percentiles of the overall price impact distribution of trades.). \protect \subref{fig:pimpact_curve} Gain in AUC from offer $p_0$ to marginal $p_{m,i}$ prices: price-sensitive (blue) and non-price-sensitive (red) traders, against minimum price difference in terms of price impact (in KL divergence). Dotted blue and dotted red lines represent 95\% CI of computed AUC differences. CI were calculated through bootstrap sampling with replacement. }
\label{fig:pimpact_figures}
\end{figure*}

%% file: Figures/fig3.tex
 \begin{figure}
  \subfloat[]{
    \centering
    	\includegraphics[width=.33\textwidth]{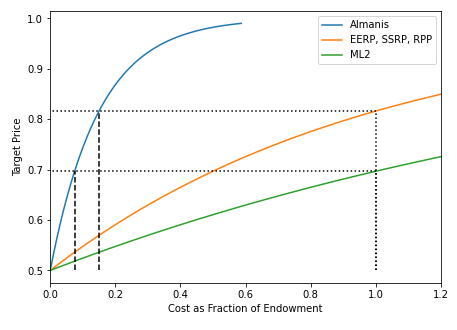}
			\label{fig:market_configuration_comparison}
    }
  \subfloat[]{
    \centering
    	\includegraphics[width=.33\textwidth]{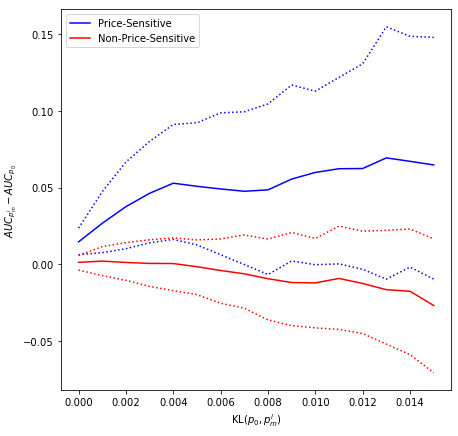}
			\label{fig:reproduction_pimpact}
    }
  \subfloat[]{
    \centering
    	\includegraphics[width=.33\textwidth]{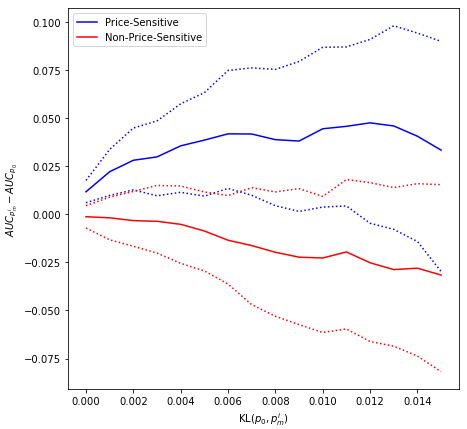}
			\label{fig:reproduction_pipact_transitive}
    }
  \caption{\protect \subref{fig:market_configuration_comparison} Marketplace configuration comparison: the X axis is the total cost as a percentage of endowment of a trade that moves the price from 0.5 to the corresponding y-axis price, using the LMSR parameter settings from the various (field) experiments (and their respective initial endowments)). Dotted black lines show the maximum price impact a trader can achieve by committing their entire endowment into a single trade in a single market (for the non-Almanis experiments). The dashed black lines are the corresponding percentage endowment that an Almanis trader would have to commit in a single trade to a single market to have the same price impact. (Note: With the LMSR, consecutive trades made by the same trader without others trading in the interim is exactly equivalent to a single aggregate trade in terms of the number of shares exchanged, the cost of those shares, and the price impact), \protect \subref{fig:reproduction_pimpact} External data: Gain in AUC from offer $p_0$ to marginal $p_{m_i}$ prices in trades by price-sensitive (blue) and non-price-sensitive (red) traders, against price impact. Dotted blue and dotted red lines represent 95\% CI of computed AUC differences. \protect \subref{fig:reproduction_pipact_transitive} External data: Gain in AUC from offer $p_0$ to marginal $p_{m,i}$ prices in trades by price-sensitive (blue) and non-price-sensitive (red) traders, against price impact, assuming transitivity of informedness between different markets. Dotted blue and dotted red lines represent 95\% CI of computed AUC differences. CI were calculated through bootstrap sampling with replacement.}
\label{fig:repro_pimpact_figures}
\end{figure}

%% file: Figures/fig4.tex
\begin{figure*}
    \centering
    \includegraphics[width = \linewidth, scale = 0.5]{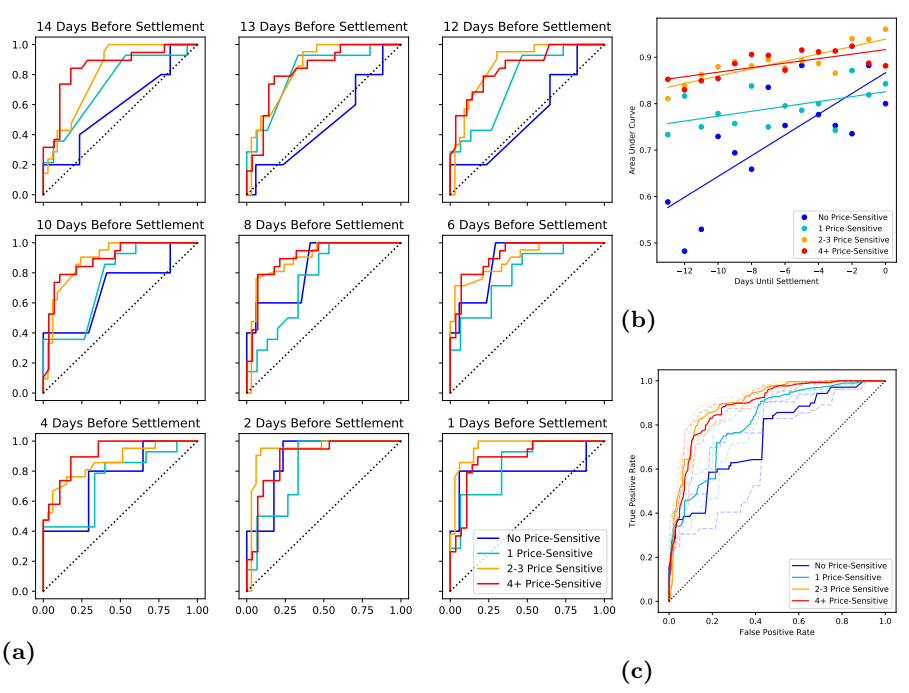}
    \caption{(a) Daily ROC curves of market forecasts from 14 days to 1 day before settlement. Markets are grouped by the number of price-sensitive traders: no price-sensitive traders (blue), 1 price-sensitive trader (teal), 2-3 price-sensitive traders (orange), 4+ price-sensitive traders (red). (b) Evolution of the AUC of the daily ROC curves of market forecasts. Markets are grouped by number of price-sensitive traders. (c) Bootstrapped temporally averaged ROC curves. The curves were made by creating 10000 sub-samples of 152 random markets (with replacement) and temporally averaging the daily ROC curves generated from the sub-sample groups through the threshold method. 
}
    \label{fig:market_price_convergence}
\end{figure*}

%% file: S_discussion.tex
\section{Discussion}

In this paper, we proposed a method for detecting traders that are price-sensitive in real-world prediction markets, and demonstrate that these price-sensitive traders are indeed informative in aggregate. Traders who are not price-sensitive had price impact consistent with noise trading. Combined these observations provide strong evidence for the Kyle model of price formation in real-world prediction markets, especially given reproduction on market data from independently operated field experiments, and prior experiment. 

It should be noted that while price-sensitive traders have a strong positive impact on the informativeness of the prices, they do so in part because the prices that these traders accept (the prices that “cause” them to trade) are markedly worse than the average market price. We cannot explain how it is that informed traders identify these bad prices, only that they demonstrably can. This also means that markets where pricing is too good (due to prior trading) are markets where our method will not work. In a market with perfect prices there are no opportunities for informed traders to repeatedly trade, and therefore there is no ability for them to generate the trade history necessary to demonstrate their price-sensitivity.

In markets where information is distributed heterogeneously and evenly (such as experiments where every market participant is given an independently drawn ball from a jar and must trade a security that pays the fraction of balls of a certain color in the jar), this means that the apparent “wisdom of the crowd” - the averaging of the information held by the market participants, is precisely because in this setup every individual participant is equally informed with independent information. Thus all traders are informed, and the market price appears as an average of their information not because this is the fundamental mechanism of price formation but rather because the Kyle model of information amplification converges to this under this specific configuration.

Additionally, in the Almanis data set the lack of consistency of price-sensitivity across markets casts into doubt the typical explanation of the motivation of market participants: that traders are drawn to a market to profit from their piece of heterogeneous information. Even people who trade in one market as if they are informed participate regularly in other markets in which they do not trade as if they are informed, making them part of the “noise-traders” in this analysis, and suggesting that they do not consistently add information in this case. A suggested motivation is that when individuals who act elsewhere as informed, act as noise-traders, they are seeking to diversify their exposure/portfolio of securities, and are happy to accept a cost for doing so (by on average worsening the price on the purchased security). Noise traders are not necessarily trading randomly because they are “irrational”. They may have various consumption needs external to the valuation of the security being traded.\cite{Admati:1988bb} There is also a possibility that some noise traders actually do have information relevant to market prices, but this information is primarily about the joint measure of the underlying outcome space that connects multiple complete event markets together, and not on the marginal measures, making them a type of arbitrageur. Future research should clarify whether this is the case.

The results of this paper highlight a major advantage of markets over alternative group forecasting systems: with a market, neither the market operator nor the end user of the price implied forecasting signal need to have information on who in the crowd of participating traders is informed. All that is necessary is that subset of the individuals participating in the marketplace are informed on the topic of interest, and the market price signal will reflect this information provided the markets are competitive. This is a direct result of the Kyle model - informed traders compete with each other for profitable trade opportunities and by doing so their information becomes fully reflected in the market price. In stark contrast, in an averaging system the choice of the average/weighting by the designer (even if dynamic) explicitly assumes or predicts which individuals will be informed (by placing greater weight on their forecasts). In a market, price quality is determined not by an average, but rather it dynamically reflects the quality of information of the informed traders, even if this is a relatively small number of individuals. This is an important distinction since simply having a large crowd will not result in good prices; only if there are informed individuals within this crowd will prices be good.

Consequently, a successful prediction marketplace should aim to engage a diverse range of individuals if it wishes to form good prices on a variety of events. Although most of the individuals in a typical market do not help consistently with reducing pricing error, a diverse crowd provides opportunities for price-sensitive traders to dissent and take a position they perceive to be profitable based on their own information and reasoning Black (1986) \cite{Black:1986et}. A sufficiently large and diverse crowd both raises the chances of there being informed individuals present on an arbitrary topic, and gives the price-sensitive traders more opportunity to engage in the market, which in turn reduces mispricing.

In conclusion, this paper presents a method the application of which improves our understanding of the process by which information enters price of a field prediction market. Foremost, our methodology detects the presence of wisdom \textit{in} the crowd (price-sensitive traders), allowing one to identify those traders that are contributing information to the market price. By doing so, our methodology helps secure an understanding on the necessary conditions for a market to produce good forecasts.

%% file: SOM_LMSR.tex
\subsection{Logarithmic Market Scoring Rule}

Logarithmic Market Scoring Rule markets are a form of subsidized market where
the market maker supplies and demands shares according to a fixed cost function.
This cost function for a simple Boolean question is expressed in equation
\ref{eq:LMSR_cost}.

\begin{equation} \label{eq:LMSR_cost}
C(q_1,q_2) = B*ln(e^{q_1/B}+e^{q_2/B}),
\end{equation} where $B$ is the liquidity constant that controls the price
impact of traders in the market, and $q_1$,$q_2$ are the quantity of shares
outstanding that pay out in the True and False settlement cases.

The marginal price of an infinitesimal unit of a share is the derivative of the
cost function with respect to the quantity of that share (equation
\ref{eq:LMSR_marginal}).

\begin{equation} \label{eq:LMSR_marginal}
p_{q_1} = \frac{e^{q_1/B}}{e^{q_1/B}+e^{q_2/B}},
\end{equation}

%% file: SOM_Methods.tex
\subsection{Methods}

The market platform that produced the data studied in this paper is called
Almanis. The Almanis website was publicly launched December 1, 2015. The
market data studied here was collected continuously from the opening of the
Almanis platform December 1, 2015 until May
2, 2017.

Recruitment of participants began slightly prior to the platform’s launch date,
with online pre-registration opening on November 1, 2015. Users were recruited
online via social media - primarily through Twitter - and through word of
mouth. Advertising on social media was done only in English, and attempted to
target users who displayed interest in forecasting on their online profiles. On
January 20, 2016 an article covering Almanis was published in the online version
of the Economist, which drew additional users to the site.

Participants were recruited through open registration, with no blanket
restrictions. As a result our participants were all self-selected,
participating voluntarily after discovering the Almanis platform. Participants
were free to join throughout the entire lifespan of the platform, until May
2, 2017. Geographic breakdown of all unique user IDs based on IP address is
shown in Table \ref{tb:demographics}. With a total 2770 unique users, our
sample is representative of the global English-speaking forecaster community.

In some cases, participants were excluded based on failure to comply with terms
and conditions. This primarily constituted of the creation of multiple
accounts, which would then trade directly with one another to transfer points
to a single account. This would place the single account high on the
leaderboard in an effort to claim a large share of that month’s payout. These
instances were handled on a case by case basis over the lifespan of the
platform. The identification process looked at the accounts that traded
consecutively most frequently, and then evaluated the performance of these
accounts relative to each other, incorporating other meta-data such as creation
time of the accounts, IP, and balance history (did one account always gain
points from the other). The decision to exclude accounts was then made by the
operations team after evaluating the available evidence.

\input{SOM_Fig_Tables/demographics}

Market topics were curated in-house by our operations team at the Dysrupt Labs
offices in Melbourne, Australia. The key subject areas of the curated markets
were: Economics, Markets, Medical, Social, Technology, and Geopolitical topics.
A complete list of the questions corresponding to each of the markets analysed
in this paper is given in an attached data file. Participants had the opportunity
to suggest potential markets through the suggestion feature of the platform.
These suggestions were vetted by the operations team and corresponding markets
were created if the topic was deemed suitable. A small number of market topics
were chosen to address the interests of partners, focusing on forecasting
macroeconomic indicators such as US Nonfarm Payroll.

Factors that were considered when creating a market topic included: adequate
forecaster interest to ensure sufficient participation, the ability to
articulate clear and unambiguous resolution criteria in advance, access
to/timeliness of resolution data, and a likely timeframe for resolution.

The specific information obtained from participant observation included the
name and email of participants upon signup. Using the crowd management system
Intercom, information on connection times, number of web sessions, browser
type, and location of users were also collected. The Almanis platform itself
recorded user trades, trade balances, and comment history. Queries from the
participants regarding difficulties with using the platform, clarifications on
market mechanics, or feedback on topics were dealt with through email by
Dysrupt Labs’ operations team.

\input{SOM_Fig_Tables/market_breakdown}

%% file: SOM_Fig_Tables/demographics.tex
\begin{table*}
\caption{Distribution of country-level location (based IP addresses) associated
with traders on the Almanis platform.}
\begin{center}
{\small
\begin{tabular}{|lr|lr|}
\hline
Country &  Unique Traders & Country & Unique Traders \\
\hline
Albania                          &        1 &	Kuwait                           &
1 \\
Anonymous Proxy                  &       52 &	Kyrgyzstan                       &
1 \\
Argentina                        &        7 &	Laos &
1 \\
Asia/Pacific Region              &        3 &	Liechtenstein                    &
1 \\
Australia                        &      306 &	Lithuania                        &
9 \\
Austria                          &        7 &	Luxembourg                       &
3 \\
Azerbaijan                       &        1 &	Macedonia                        &
1 \\
Bangladesh                       &        2 &	Malaysia                         &
2 \\
Belgium                          &       12 &	Malta                            &        2 \\
Botswana                         &        1 &	Mexico                           &       11 \\
Brazil                           &       24 &	Morocco                          &        2 \\
Brunei Darussalam                &        1 &	Nepal                            &        3 \\
Bulgaria                         &        3 &	Netherlands                      &       23 \\
Canada                           &      105 &	New Zealand                      &       14 \\
Cayman Islands                   &        1 &	Nigeria                          &        7 \\
Chile                            &        2 &	Norway                           &        9 \\
China                            &        4 &	Pakistan                         &        7 \\
Colombia                         &        8 &	Peru                             &        1 \\
Costa Rica                       &        2 &	Philippines                      &       22 \\
Cyprus                           &        6 &	Poland                           &       18 \\
Czech Republic                   &        6 &	Portugal                         &       11 \\
Denmark                          &        5 &	Puerto Rico                      &        1 \\
Dominican Republic               &        1 &	Qatar                            &        2 \\
Ecuador                          &        1 &	Reunion                          &        1 \\
Estonia                          &        7 &	Romania                          &       11 \\
Europe                           &        4 &	Russian Federation               &       15 \\
Finland                          &       12 &	Saudi Arabia                     &        2 \\
France                           &       23 &	Senegal                          &        1 \\
Georgia                          &        1 &	Serbia                           &       13 \\
Germany                          &       42 &	Singapore                        &       31 \\
Ghana                            &        2 &	Slovakia                         &        4 \\
Greece                           &        8 &	Slovenia                         &        4 \\
Guinea                           &        3 &	South Africa                     &       15 \\
Hong Kong                        &       16 &	Spain                            &       21 \\
Hungary                          &        3 &	Sweden                           &       33 \\
Iceland                          &        1 &	Switzerland                      &       17 \\
India                            &      146 &	Taiwan                           &        1 \\
Indonesia                        &        6 &	Tanzania     &        1 \\
Iran        &        1 &	Thailand                         &        2 \\
Ireland                          &       12 &	Tunisia                          &        1 \\
Isle of Man                      &        2 &	Turkey                           &       10 \\
Israel                           &       21 &	Ukraine                          &        6 \\
Italy                            &       26 &	United Arab Emirates             &        8 \\
Japan                            &        6 &	United Kingdom                   &      948 \\
Jersey                           &        1 &	United States                    &      538 \\
Jordan                           &        3 &	Uruguay                          &        2 \\
Kazakhstan                       &        2 &	Venezuela                        &        3 \\
Kenya                            &        5 &	Vietnam                          & 
4 \\
Korea, Republic of               &        4 &	Zambia                           &
1 \\

\hline
\end{tabular}
}
\end{center}
\label{tb:demographics}
\end{table*}

%% file: SOM_Fig_Tables/market_breakdown.tex
\begin{landscape}
\begin{table}
\begin{tabular}{llrlrlrrrrr}
\toprule
                  &               & \multicolumn{2}{l}{$\Delta$AUC pim vs p0 PS} & \multicolumn{2}{l}{$\Delta$AUC pim vs p0 nonPS} & $\Delta\Delta$AUC & \multicolumn{2}{l}{KLdiv [nats]} & \multicolumn{2}{l}{Volume} \\
                  &               &            Value &                95\%CI &               Value &                 95\%CI &      Value &       median &      IQR &     PS &  nonPS \\
\midrule
Days to EOS & [0.0, 5.0) &          0.03941 &   [0.01964, 0.05919] &            -0.00257 &   [-0.01065, 0.00551] &  0.04199 &      0.00872 &  0.05156 &   1838 &   7456 \\
                  & [5.0, 10.0) &          0.02324 &  [-0.00218, 0.04866] &            -0.00288 &   [-0.01517, 0.00942] &  0.02612 &      0.00872 &  0.05045 &   1099 &   3476 \\
                  & [10.0, 14.0) &          0.02195 &  [-0.00586, 0.04977] &            -0.00642 &   [-0.02276, 0.00992] &  0.02837 &      0.00715 &  0.04601 &    819 &   2361 \\
                  & [14.0, 30.0) &          0.04091 &    [0.01931, 0.0625] &            -0.01104 &   [-0.02322, 0.00115] &  0.05194 &      0.00988 &  0.05434 &   1933 &   4817 \\
                  & [30.0, inf) &          0.03760 &   [0.02437, 0.05083] &            -0.01192 &  [-0.02083, -0.00301] &  0.04952 &      0.01808 &  0.07578 &   8033 &  15639 \\
Number of Traders & (0.999, 33.0] &          0.03747 &   [0.01971, 0.05524] &             0.00307 &   [-0.00482, 0.01097] &  0.03440 &      0.01219 &  0.05762 &   3190 &  13357 \\
                  & (33.0, 60.0] &          0.03546 &   [0.02078, 0.05014] &            -0.00969 &    [-0.019, -0.00038] &  0.04515 &      0.01262 &  0.06295 &   4418 &  10708 \\
                  & (60.0, 217.0] &          0.03721 &   [0.02269, 0.05174] &            -0.01927 &  [-0.02856, -0.00998] &  0.05648 &      0.01401 &  0.07200 &   6114 &   9684 \\
\bottomrule
\end{tabular}
\caption{Breakdown of dataset binned by Days to End of Series (EOS) and the Number of Traders participating in each market (bin limits correspond to tertiles of the distribution of number of traders). The first two major column shows Differences in Area Under Curve ($\Delta$AUC) differences between final marginal price (pim) and initial offer price (p0) for Price-Sensitive (PS) and non-Price-Sensitive (nonPS) traders. The third major column shows difference in the difference in AUC between PS and nonPS traders ($\Delta\Delta$AUC), as well as the KL divergence (KLdiv) for final marginal and initial prices. The final major column shows the trade volume (number of unique trades) by PS and nonPS traders.}
\end{table}
\end{landscape}

%% file: SOM_ROC.tex
\subsection{Receiver Operating Characteristic Curves}

\begin{figure}[h]
\centering
\includegraphics[width=0.75\linewidth]{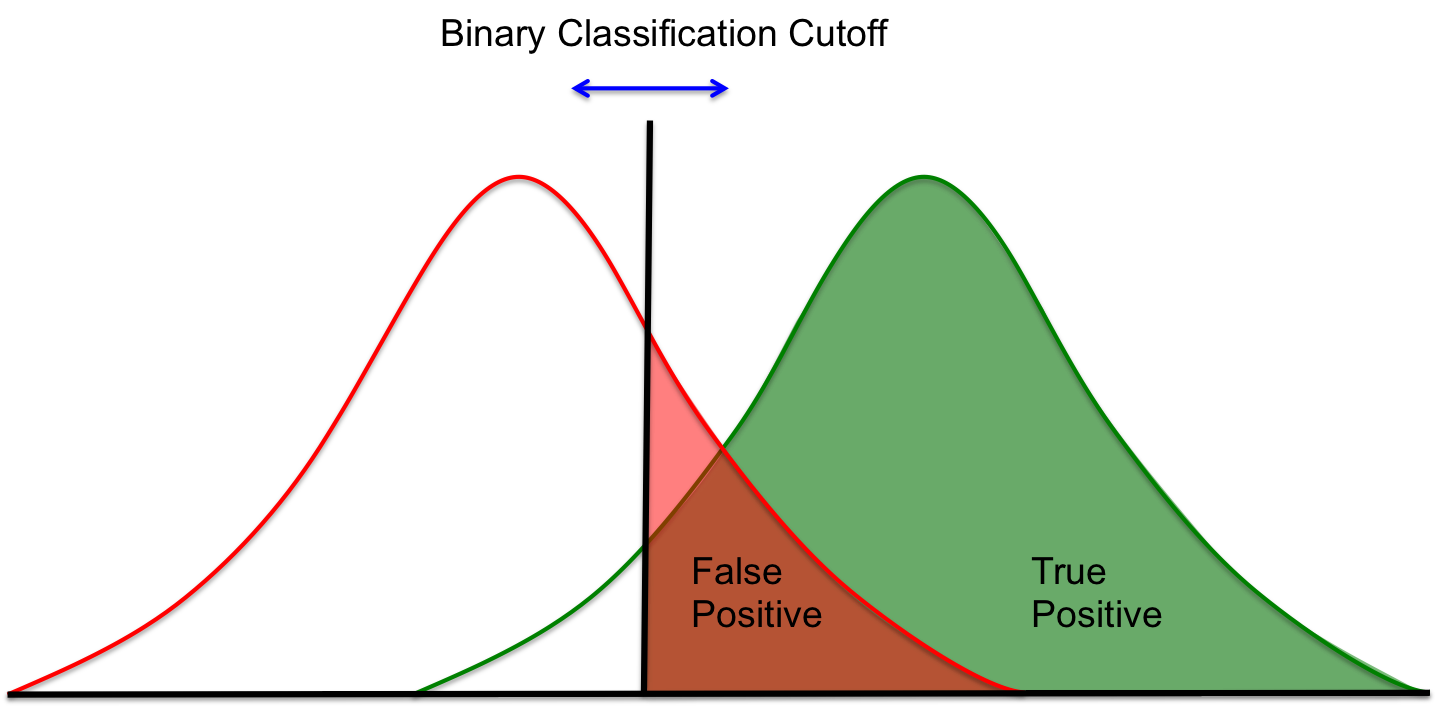}
\caption{The green curve represents the distribution of outputs of the system
under a condition positive, and the red curve represents the distribution of
outputs of the system under a condition negative. The shaded red and green
areas represent the true positive and false positive regions of the curve for
the arbitrary binary classification cutoff illustrated.}
\label{fig:ROC_curve_explained}
\end{figure}

In order to compare the different groups of markets, Receiver Operating
Characteristic (ROC) curves were used. ROC curves are useful to demonstrate the
forecasting ability of a binary classifier system. They plot the True Positive
Rate (TPR) of a classification system against its False Positive Rate (FPR).
The curve is generated through the definition of a binary classification cutoff
and shifting it through the spectrum of outputs of the system. This process is
illustrated in Fig \ref{fig:ROC_curve_explained}.

In the context of a prediction market, this means the following. Suppose there
is a market forecasting the question ``Will Donald Trump Win the 2016
Presidential Election?'', and the current price of the market is indicating a
0.4 likelihood. Does this mean that the market is forecasting Trump will lose?
Common sense would dictate that if the price indicates a likelihood less than
0.5, the market is forecasting that Trump will lose most of the time, so a
price of 0.4 is forecasting that Trump is expected to lose. An ROC curve is
constructed by shifting this likelihood threshold of 0.5 across the unit
interval in order to observe the changes in output characteristics of the
system. By redefining the threshold to 0.2, for example, it is possible to
observe how the market's classification ability changes (how many more
true positive events do you identify at the cost of how many more false alarms)
when any price above 0.2 indicates Trump will win. For the analysis performed
in this paper, the discrimination threshold was varied in 0.05 increments to
produce the curves.

The area under a ROC curve is the probability that the system will rank a
positive event higher than a negative one. Different forecasting systems can be
compared via the area under their respective curves, with greater area
indicating greater forecasting power. The AUC can be computed as:
\begin{equation} \label{eq:ROC_AUC}
A_{ROC} = \frac{1}{P*N} \int_{0}^{1} T dF,
\end{equation} where $P$ and $N$ are the number of positive and negative
instances observed, $T$ is the number of true positives as a function of the
number of false positives, and $F$ is the number of false positives.

%% file: SOM_TLS.tex
\subsection{Total Least Squares}

The following section gives a brief description of the problem posed by the potential budget constraints faced by traders in the external replication dataset, and the means of accounting for this challenge via total least squares. For a rigorous overview of TLS, please see the the references \cite{markovsky2007overview}. 

The price-sensitivity regression performed in this paper has the following form:

\begin{equation} \label{eq:price_sense_estimation_som}
	\Delta p^{R}_{i}(t_i+1)=\alpha_i+\beta_i (p_0 (t_i+1)-p_{m,i}(t_i)) + \epsilon(t_i)
\end{equation}

It assumes that $p_{m,i}(t_i)$ is a good approximation of the belief of the trader at time $t_i$. If, for any reason (budget constraints, risk aversion, etc.) this is not true, the indepdent variable of the regression $(p_0 (t_i+1)-p_{m,i}(t_i))$ will be estimated with error, biasing the estimated $\beta_i$ downwards. To see this take the following simple data generating process:

\begin{equation}
    y = b x + e
\end{equation}
For $e \sim N(m_1, s_1)$. Assume that $x$ is observed with error: 
\begin{equation}
    \hat{x} = x + f
\end{equation}
For $f \sim N(m_2, s_2)$, $cov(y, f) = 0$. If one replaces $\hat{x}$ with $x$ when performing OLS, the equation being estimated becomes:

\begin{equation}
    y = c \hat{x} + v = c x + c f + v 
\end{equation}

Note that $c$ will be a biased estimate of $b$, with the bias depending on the level of noise $s_2$ of $f$. Instead, one must specify a regression model assuming errors in the independent variable:

\begin{equation}
    y = c (x + f) + v 
\end{equation}

\begin{figure}[h]
\centering
\includegraphics[width=0.75\linewidth]{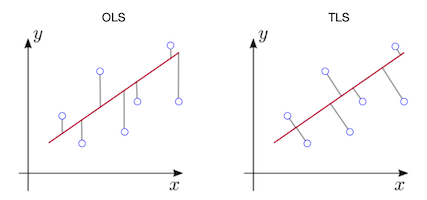}
\caption{OLS minizes error in the dependent variable. TLS minimizes error orthogonal to the line of fit.}
\label{fig:OLS_TLS}
\end{figure}

And estimate c using Total Least Squares. TLS is performed via the singular value decomposition (SVD) of the data matrix $[y; \hat{x}] = USV^{\top}$, and rescaling the eigenvector $U[:,-1]$ corresponding to the smallest singular value to have unit value for the dependent variable $y$. This gives the equation for a hyperplane which minimizes the orthogonal error to the data points, as shown in Figure \ref{fig:OLS_TLS}.

%% file: SOM_Truncation.tex
\subsection{Trade Frequency Based Truncation}
\label{subsec:truncation}

A significant number of market time series had a long “tail” of low trade
activity before settlement. This is frequently due to the fact that a critical
piece of information determining the outcome of the question becomes public
knowledge. The outcome of the question was therefore considered a foregone
conclusion, but since the question had not yet been settled, the market
remained open for trading. As a result, these markets simply reflect the current
reality, rather than a meaningful prediction, making them incomparable with
markets where the outcome is difficult to predict up to settlement. These
market time series were therefore truncated before the analysis to remove
periods of low trade activity.

The methodology of truncation was as follows: if the overall trade frequency of
the market fell below a certain value (12 hours per trade in the analysis
presented), the trade closest to settlement was dropped. This process was
repeated until the market either attained the desired average trade frequency or
dropped below 25 total trades, in which case the market was dropped from the
analysis. No further selection criteria were applied to the market data.